# High-performance $Ba_{1-x}K_xFe_2As_2$ superconducting joints for persistent current operation

Yanchang Zhu[1,2,3], Dongliang Wang[1,2,3*], Qiqi Wang[5], Wenwen Guo[1,2,], Cong Liu[1,2,3], He Huang[1,2,3], Hongjun Ma[5], Fang Liu[5], Huajun Liu[5], and Yanwei Ma[1,2,3,4*]

[1] Key Laboratory of Applied Superconductivity, Institute of Electrical Engineering, Chinese Academy of Sciences, Beijing 100190, People's Republic of China

[2] University of Chinese Academy of Sciences, Beijing 100049, People's Republic of China

[3] Institute of Electrical Engineering and Advanced Electromagnetic Drive Technology, Qilu Zhongke, Jinan, 250013, People's Republic of China

[4] School of Materials Science and Engineering, Zhengzhou University, Zhengzhou, 450001, People's Republic of China

[5] Institute of Plasma Physics, Chinese Academy of Sciences, Hefei 230031, People's Republic of China

**Abstract**

Superconducting joints are one of the key technologies to make $Ba_{1-x}K_xFe_2As_2$ (Ba-122) superconducting wires or tapes for high-field applications. Herein, superconducting joints were fabricated by a simple cold-pressing method, and the joint resistance of the iron-based superconducting joint was estimated for the first time. The superconducting properties, microstructures, and elements distribution in the joint regions were investigated. At 4.2 K and 10 T, a transport critical current $I_c$ of 105 A for the joint was obtained, and the critical current ratio (CCR= $I_c^{joint}/I_c^{tape}$) of the joint was 94.6％. On the other hand, the joint show very low joint resistance of 2.7×$10^{-13}$ Ω in self-field at 4.2 K. Among iron-based superconductors (IBS), this work is the first to successfully realize a superconducting joint with such high CCR and low joint resistance. This work shows great potential to apply Ba-122 in a range of practical applications, where superconducting joints are essential.

---
*Author to whom correspondence should be addressed: E-mail: dongliangwang@mail.iee.ac.cn or ywma@mail.iee.ac.cn

1. Introduction

The K doped '122' type (AeFe$_2$As$_2$, Ae= alkali or alkali earth elements) iron-based superconductors are considered to be one of the most promising superconducting materials for the fabrication of high-field magnets , due to their high superconducting transition temperatures ($T_c$) up to 38 K [1], very high upper critical field ($H_{c2}$) above 100 T [2, 3], small anisotropy with $\gamma = \| ab/H_{c2} \| c$ being about 1.5-2 and large critical current density ($J_c$) over $10^6$ A·cm$^{-2}$ in thin film [4-7]. Significant process toward high-performance iron-based superconductors has been made over recent years [8-12]. For instance, Ba-122 tapes can reach large transport $J_c$ of 1.5×10$^5$ A·cm$^{-2}$ at 10 T and 4.2 K, and still maintains 10$^5$ A·cm$^{-2}$ at 14 T and 4.2 K in short sample [13], and the 100-m class Sr$_{1-x}$K$_x$Fe$_2$As$_2$ (Sr-122) tapes were fabricated in 2016 [14]. All these prominent characters of iron-based superconductors are favorable for high field applications.

However, high-performance superconducting joints will be the key point to employ iron-based superconductors in future MRI systems. This is because MRI magnets are usually operated in the persistent mode to produce an ultrastable magnetic field with a decay rate of < 0.1 ppm h$^{-1}$ [15]. Furthermore, the first iron-based superconducting joint was fabricated by the hot-pressing (HP) method [16]. The achieved critical current ratio (CCR) was 35% at 4.2 K and 10 T and the joint resistance was estimated to be below 10$^{-9}$ Ω [16]. However, they found that the transport properties of iron-based superconducting joints are affected mainly by the superconducting phases and microstructures in the connection areas. Though, such cracks and K loss were nearly eliminated by optimizing HP condition, resulting in CCR rising to 63% [17]. S Imai *et al* fabricated superconducting joints between BaK-122 tapes using a cold-press method, and a CCR of ~90% was achieved at 4.2 K and 3.5 T [18]. Nonetheless, they find that Ba and O are rich (K, Fe, and As are poor) along the joint interface, and still they haven't measured the exact resistance of the superconducting joint.

In the paper, we prepared Ba-122 superconducting joints via a simple cold-pressing method, which is considered to be preferable for the practical use compared with the HP method. Samples of the Ba-122 joints were connected by the cold-pressing method. The problem of inhomogeneous along the joint interface has been solved and the macro-cracks in the joint region are well controlled. The critical current $I_c$ of the best joint was 105 A at 4.2 K and 10 T. The resistance of the iron-based superconducting joint was first measured by the field-decay measurement. The resistance of the joint was $2.7 \times 10^{-13}$ Ω in self-field at 4.2 K.

## 2. Experimental details

The Ba-122 tapes were fabricated by the ex-situ PIT method. Firstly, Ba filings, K pieces, As powders were loaded into Nb tube under argon atmosphere for 10 hours, and then sintered at 900 °C for 35 hours. The sintered precursor was ground into fine powder in an agate mortar in an argon atmosphere. Then the fine powder was packed into Ag tubes (OD 8 mm and ID 5 mm). These tubes were finally drawn into wires of 1.9 mm in diameter and then rolled into tapes of thickness 0.4 mm. Finally, the tapes were sealed into quartz tubes and heat treated at 880 °C for 0.5 h in an Ar atmosphere.

Figure 1 shows the schematic of Ba-122 superconducting joints. Firstly, as is shown in figure 1(a), the sheath materials were partially peeled off using mechanical polishing until the Ba-122 core was exposed. After removing the metallic sheath, exposed core of the tapes were aligned and made to face each other using the Ag foil. Then to make close contact with the tape core, uniaxially pressure was applied, shown in figure 2(c). Table 1 summarizes the details of joint samples prepared at different pressures. Finally, the joint was heat treated at 880°C for 0.5 h in an Ar atmosphere.

To determine joint resistance, a field-decay measurement was carried out. The field-decay measurement setup is shown in figure 2. A Hall sensor with 0.1 Gs sensitivity was installed at predefined location to record the magnetic field generated by the coil during the test. The temperature sensor was also installed on the top of the joint to monitor the temperature of the field-decay test. Following the fabrication steps illustrated in figure 1, a Ba-122 superconducting joint was made between the two terminals of the coil to form a closed-loop circuit, as shown in figure 2(b). Then the whole closed-loop include the skeleton was heat treated at 880 °C for 0.5 h in an Ar atmosphere. In the field-decay measurement, the background field was excited when the closed-loop coil was not superconducting. Subsequently, we made the closed-loop superconducting, and the background field was slowly reduced to zero.

We used the X-ray Computed Tomography (XCT, Model: GE Microme|x) to analyze the cracks distribution of the connection area of the joint samples. Microstructures in joint area were observed by Electron Probe Microanalysis (EPMA, Model: JEOL JXA8230). The temperature dependence of the resistivity at 0 T was carried out using a four-probe method by PPMS (Model: PPMS-9). Transport critical currents ($I_c$) of the samples were measured at 4.2 K using a standard four-probe technique, with a criterion of 1 $\mu$V cm$^{-1}$. The magnetic field dependence of transport $I_c$ values for all samples were evaluated at the High Field Laboratory for Superconducting Materials (HFLSM) in Sendai or Institute of Plasma Physics, CAS in Hefei.

## 3. Results and discussion

3.1 Transport properties of the superconducting joints

To evaluate the current-carrying capacity of the jointed Ba-122 conductor, firstly, we carried out $I_c$ measurements, as shown in figure 3. The data shows in figure 3(a) present the transport properties of Ba-122 superconducting joints at 4.2 K and magnetic fields from 4 T to 14 T. The $I_c^{joint}$ and $I_c^{tape}$ curve shows a weak magnetic field dependence, which indicates that the superconducting joints fabricated by this method are favorable for high field applications

The critical current and the critical current ratio (CCR) at 10 T and 4.2 K of the joint samples with different pressure are shown in figure 3(b). It is found that the critical current and the CCR significantly increases with the pressure increases from 0.69 to 2.3 GPa. However, when the pressure comes to 2.3~3.68 GPa, it seems that the critical current and the CCR are almost the same at 4.2 K and 10 T. The highest critical current, which is obtained at 3.68 GPa (CP80), is up to 105 A at 4.2 K and 10 T, and the critical current is 94.6%. The value of which is almost the same as that of the unjointed tapes.

3.2 Microstructure of the superconducting joints

In order to investigate the relationship between the micro-cracks and the transport properties of the joint samples, the longitudinal cross sections of Ba-122 superconducting joints were examined by SEM, as shown in figure 4. Figure 4(a)~(f) show the longitudinal cross sections of Ba-122 superconducting joints pressed at 0.69, 1.38, 1.84, 2.3, 3.22 and 3.68 GPa. In figure 4 (a), we can see that there are many holes and micro-cracks in the connection area. Meanwhile, as the pressure increases, the number of holes and micro-cracks around the connection area decreases. When pressure increases to 1.84 GPa (CP40), as shown in figure 4 (c), there is almost no obvious micro crack in the connection area. This indicates that the micro-cracks in the whole joint section significantly decrease with increasing pressure. So that we can see the CCR of CP50 to CP80 is better than that of CP15 to CP40.

The study on the transport properties of superconducting twin films shows that iron-based superconductors have the problem of weak grain boundary connection similar to copper oxide superconductors [19]. Though the former has less stringent requirements on grain boundary angle than the latter, the texture of superconducting core is still the key to realize the high transmission performance of iron-based superconductors. Figure 5 shows the EBSD characterization results of the connection area of joint CP80, and the observation surface is the longitudinal section of the joint. Figure 5(a) gives the inverse pole figure (IPF) maps of joint sample CP80. We can find that the main colors of the connection area are green and blue, which shows significant orientation. This means that there is an obvious c-axis texture in the superconducting phase in the connection area, which is beneficial to the transmission of superconducting current. However, figure 5(b) shows that there are mainly small grains in the connection area, which also shows that the grains in the connection area have not fully grown.

On the basis of scanning electron microscopy and electron probe microanalysis, it was said that micro-cracks and loss of K at the connection area will degrade the CCR performance of joints. We conducted SEM/EPMA image to investigate the influence of CP process on the microstructure of the connection area of the joint samples. An SEM image of the connection area of joint sample CP80 is shown in figure 6. We can see that the connection area has healed well, and the joint interface is almost invisible. On the other hand, the EPMA image reveals that there is almost no element loss along the joint interface. This shows that the microstructure in the connection area of the joint prepared by cold-pressing can reach the same level as that of the joint prepared by hot-pressing.

3.3 Macrostructure of the superconducting joints

Figure 7 shows the XCT images of joints made by hot-pressing (HP) method and cold-pressing method. The XCT images were taken from the longitudinal side of the joint samples. As is shown in the picture that there are many macro-cracks in the joint region of hot-pressing joint sample, however for cold-pressing joint sample there almost have no macro-cracks in the joining area. Our previous work has demonstrated that macro-cracks are harmful to the transport performance of the joint samples, and macro-cracks were almost unavoidable [16, 17]. In this study, in order to solve this problem, we replace the hot-pressing method by cold-pressing method. Obviously, the macro-crack has been well controlled, and the result is in line with our expectations. In our opinion, that silver is very soft at high temperature, so hot pressing at high temperature will bite silver into the crack, which will enlarge the crack in the joint connection area. There is no such problem as cold pressing.

3.4 Characterization of the superconducting joints with the field decay method

Finally, to evaluate the performance of the joint in an actual MRI magnet system scenario, that is, in the persistent mode, the field-decay technique was employed. This method is extremely sensitive to decay in the transport current and resistance of the joint. The decay behavior normally has two stages as shown in figure 8. The first stage shows exponentially decreasing current with a high decay rate, due to lower *n*-value and small difference between the induced current an $I_c$ of the closed-loop [20, 21]. In second stage, the decay is quite slow and mainly depends on the joint resistance. During entire field-decay measurement period, the joint was exposed to only self-field at 4.2 K. The magnetic field was allowed to stabilize for about 1000 s, and at this point, was considered to be the initial field $B_0$ at time $t_0$ for the resistance estimation. The joint resistance was estimated from temporal decay of magnetic field in the time constant of *L-R* circuit:

$$B=B_0\,e^{-(R/L)},$$

Where *B* is the field at time t, $B_0$ is the magnetic field at time $t_0$, *L* is the inductance of the closed-loop coil. The estimated joint resistance according to the *L-R* circuit decay equation was $2.7\times10^{-13}\,\Omega$ in self-field at 4.2 K. The measured joint resistance meets the application requirements of such joints in persistent mode MRI magnets. This is very exciting news for the practical application of iron-based superconductors in the future.

According to our results, it is clear that the transport properties of Ba-122 superconducting joints could be significantly improved by cold-pressing. Firstly, in this work, the problem of inhomogeneous distribution of elements along the joint interface has been well controlled. As is known to us, the elements along the joint interface are easy to lose during the fabrication of superconducting joint. This will result in a wide superconducting transition temperature $T_c$ [16, 18]. Secondly, we find that the micro-cracks and holes along the joint interface could be controlled by increase the pressure during the CP process. In our previous result, the micro-cracks could block he transmission of current at the joint interface. This will lead to poor superconducting properties. Most importantly, we find that there almost have no macro-cracks in the joint region compared to the HP joint. In previous studies, macro-cracks were almost unavoidable. This may be one of the key points for the fine superconducting performance of CP joint. However, in this case, we find that there are many small grains in the connection area, which means that the grains in the connection area have not fully grown. In the future, further increase of the transport for the joint is expected by trimming heat treatment parameters to make the grains in the connection area fully grown.

4. Conclusions

In conclusion, we fabricated the superconducting joints between Ba-122 tapes by the cold-pressing method, and estimated the iron-based superconducting joint resistance by field-decay method for the first time. Unprecedentedly, the joint shows almost identical current-carrying capacity compared to the bare tape from 4 to 14 T at 4.2 K, joint resistance of $2.7 \times 10^{-13}$ Ω in self-field at 4.2 K. Our microstructure SEM, EPMA and EBSD examination revealed a very small crack, few K loss and less texture in the joint cross-section area. The macrostructure XCT examination revealed there are almost no macro-cracks in the joint cross-section area. However, these artifacts did not affect the performance of the joint. There is no doubt that this result will be very conducive to the practical application of iron-based superconductors in the future.

**Acknowledgments**

The authors would like to thank Prof. S. Awaji at Sendai for $I_c$-B measurement. This work is partially supported by the National Key R&D Program of China (Grant Nos. 2018YFA0704200 and 2017YFE0129500), the National Natural Science Foundation of China (Grant Nos. 51861135311, U1832213 and 51721005), the Strategic Priority Research Program of Chinese Academy of Sciences (Grant No. XDB25000000), Key Research Program of Frontier Sciences of Chinese Academy of Sciences (QYZDJ-SSW-JSC026), the International Partnership Program of Chinese Academy of Sciences (Grant No. 182111KYSB20160014).

Table 1. Details of joint samples prepared at different pressures.

| Joint name | CP15 | CP20 | CP25 | CP30 | CP35 | CP40 | CP50 | CP60 | CP70 | CP80 |
|---|---|---|---|---|---|---|---|---|---|---|
| Pressure(GPa) | 0.69 | 0.92 | 1.15 | 1.38 | 1.61 | 1.84 | 2.3 | 2.76 | 3.22 | 3.68 |

**Captions**

Figure 1.A schematic picture of joining process: (a) peeling off the sheath metal; (b) docking superconducting window；（c）wrapping in Ag foil and cold-pressing; (d) sintering.

Figure 2. (a)Ba-122 closed-loop coil and its field-decay test setup; (b) Ba-122 closed-loop coil.

Figure 3.(a) Magnetic field dependence of transport critical current at 4.2 K for joints and tapes. (b) Critical current and CCR versus pressure curve.

Figure 4.SEM image of longitudinal cross section of Ba-122 superconducting joints.

Figure 5. Grain orientation analysis of joint CP80 in Rolling direction (RD). (a) IPF map of joint CP80. (b) Grain size distribution by quantity of joint CP80.

Figure 6.SEM and EPMA mapping image of the joint CP80.

Figure 7. XCT images of the HP joint and CP joint.

Figure 8.Time decay curve of the captured magnetic (y-axis field is the induced field in the closed-loop coil).

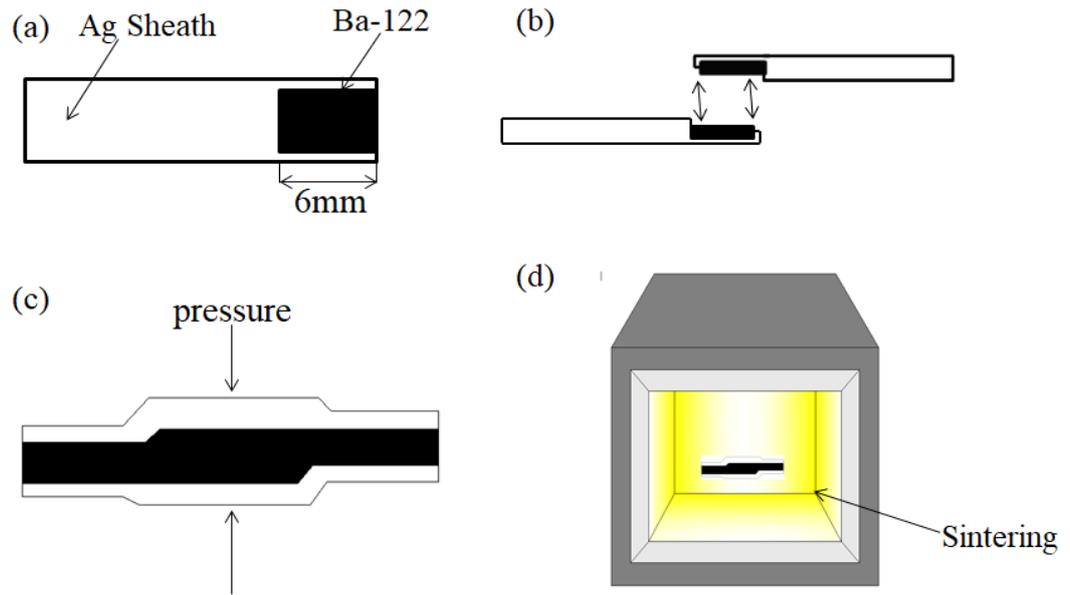

Figure 1. Zhu et al

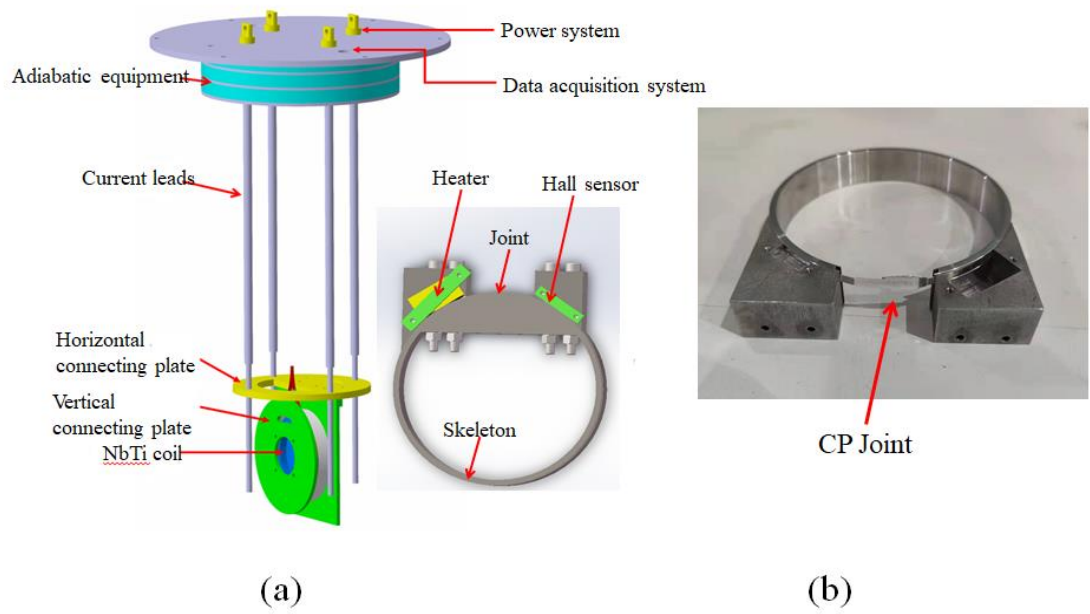

Figure 2. Zhu et al

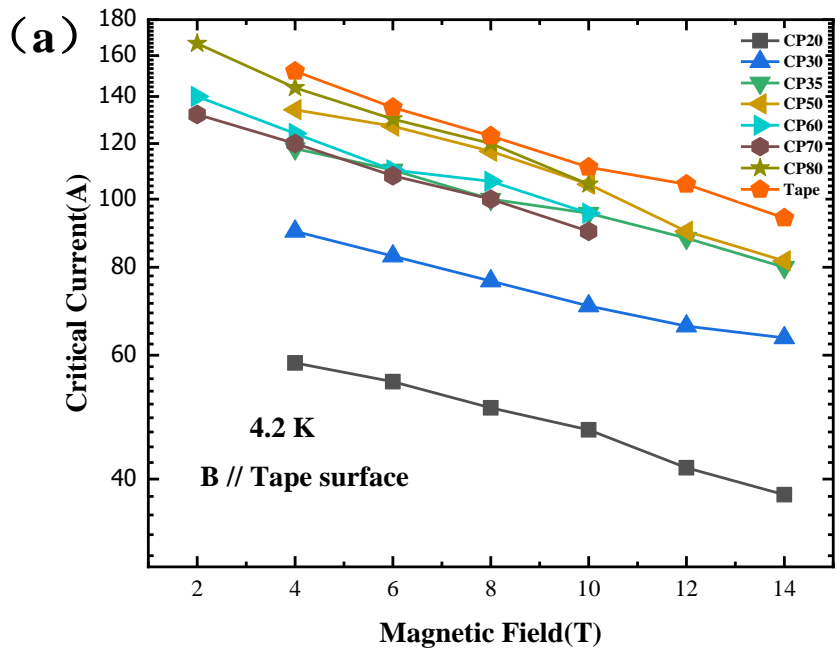

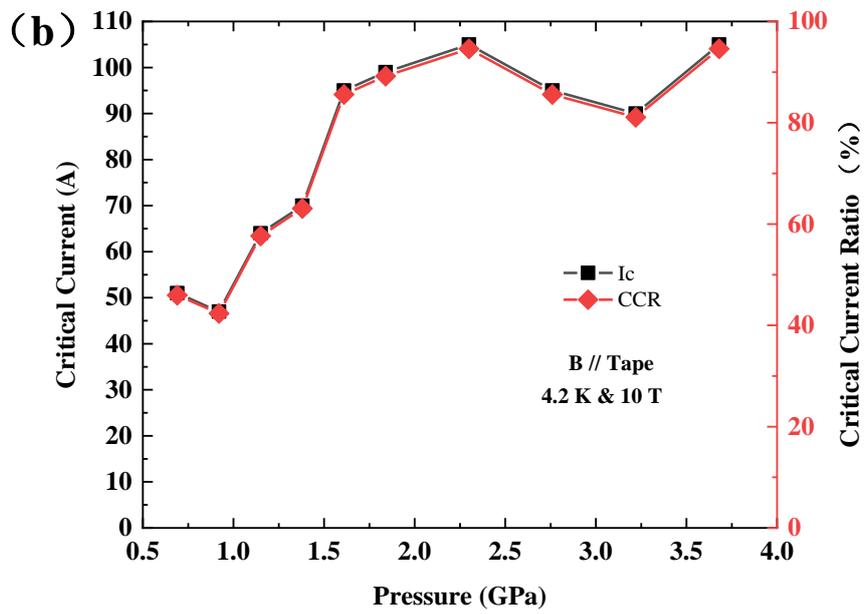

Figure 3. Zhu et al

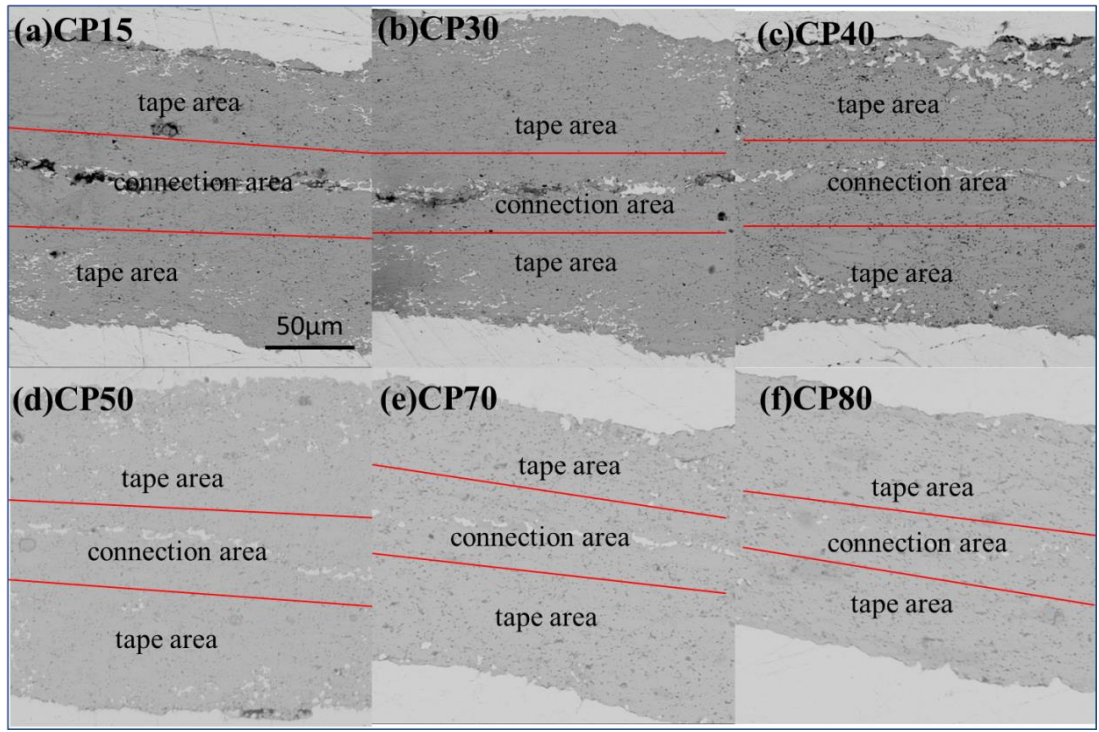

Figure 4. Zhu et al

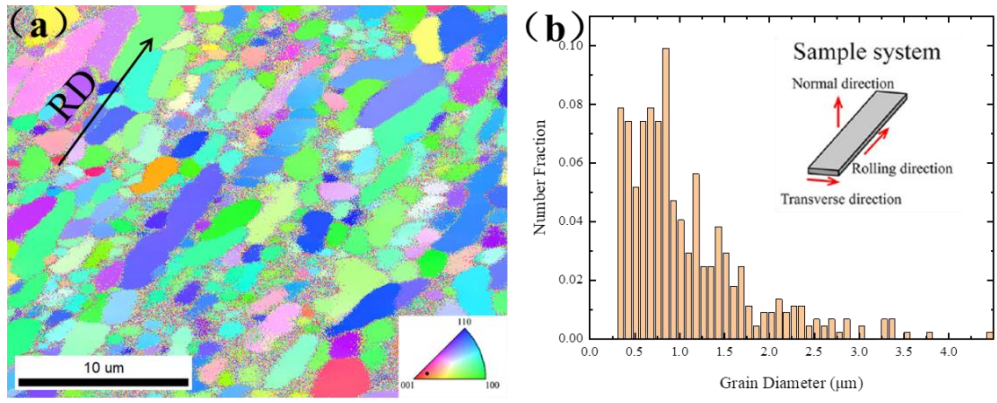

Figure 5. Zhu et al

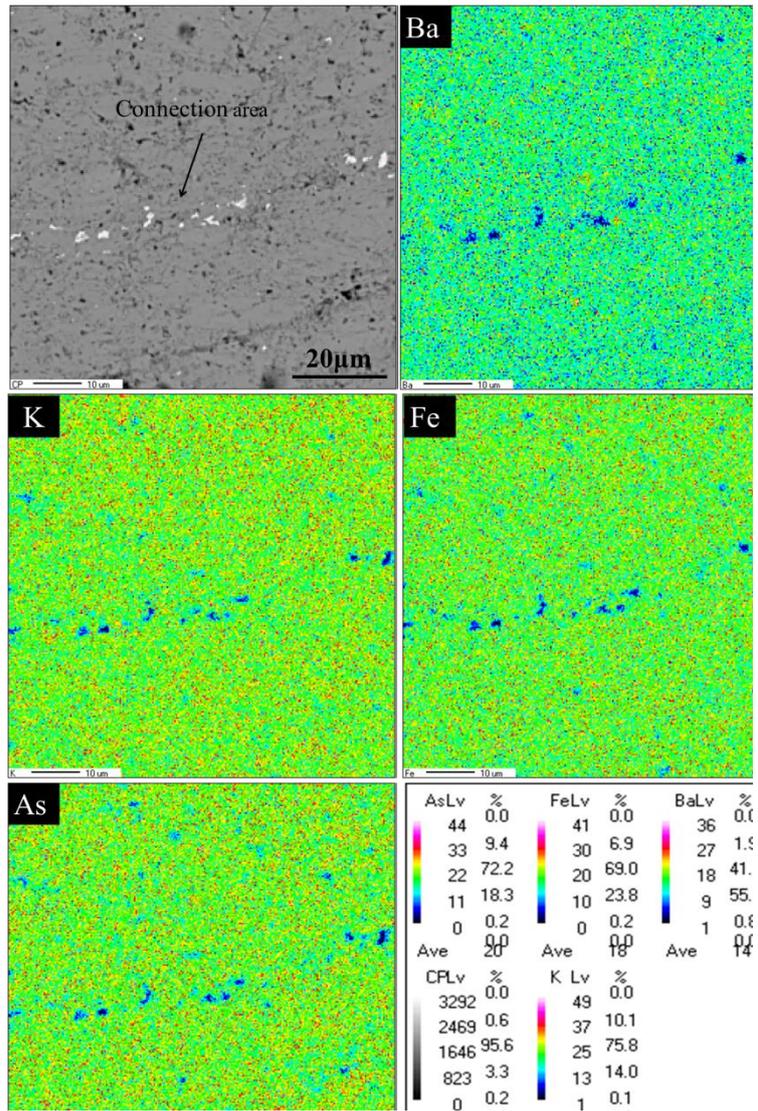

Figure 6. Zhu et al

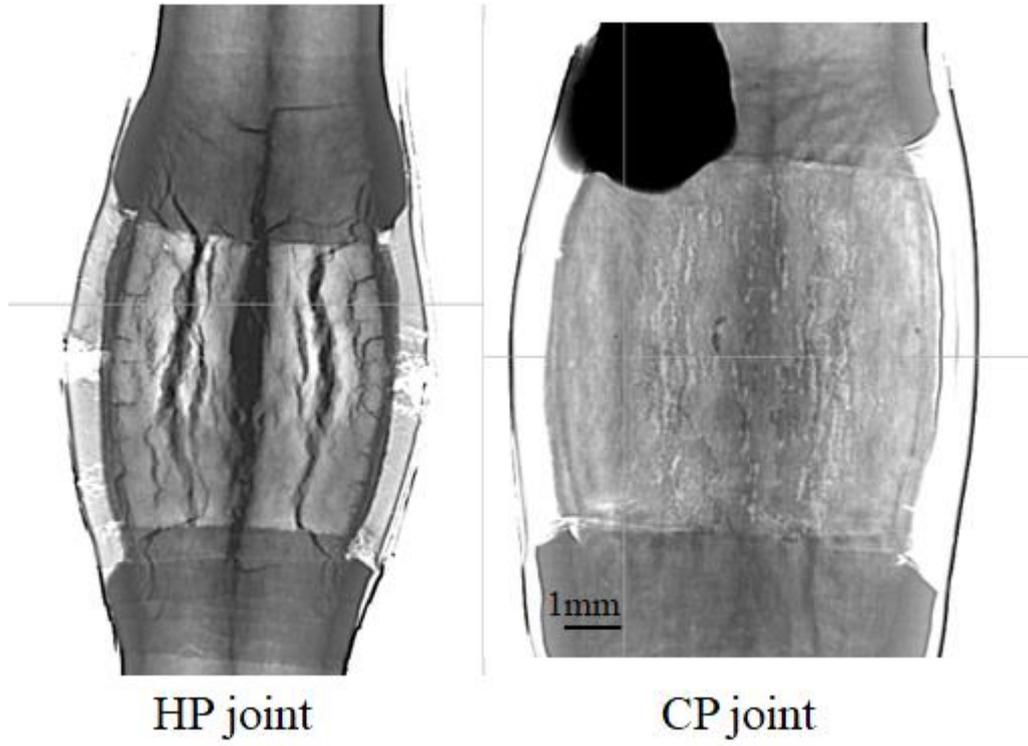

Figure7. Zhu et al

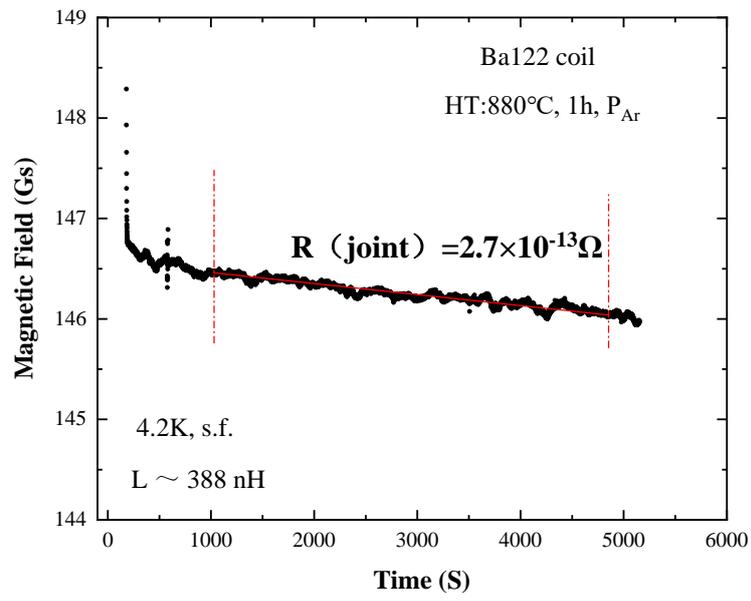

Figure 8. Zhu et al